\documentclass[11pt,journal,twocolumn]{IEEEtran}

\usepackage{cite}

\usepackage{graphicx,color,epsfig,rotating,subfigure}
\usepackage{amsfonts,amsmath,amssymb}
\usepackage{algorithm,algorithmic}
\usepackage{algorithm,algorithmic}
\usepackage{url}
\usepackage[stable]{footmisc}

\long\def\comment#1{}

\newfont{\bbb}{msbm10 scaled 700}

\newfont{\bb}{msbm10 scaled 1100}

\newcommand{\RR}{\mbox{\bb R}}

\newcommand{\Rm}{{\bf R}}

\newcommand{\Xm}{{\bf X}}

\newcommand{\Ac}{{\cal A}}
\newcommand{\Bc}{{\cal B}}

\newcommand{\Ec}{{\cal E}}
\newcommand{\Fc}{{\cal F}}
\newcommand{\Gc}{{\cal G}}
\newcommand{\Hc}{{\cal H}}
\newcommand{\Ic}{{\cal I}}

\newcommand{\Mc}{{\cal M}}
\newcommand{\Nc}{{\cal N}}

\newcommand{\Pc}{{\cal P}}

\newcommand{\Uc}{{\cal U}}

\newcommand{\Omegam}{\hbox{\boldmath$\Omega$}}

\newtheorem{definition}{Definition}
\newtheorem{theorem}{Theorem}
\newtheorem{lemma}{Lemma}

\newtheorem{problem}{Problem}

\begin{document}

\title{FemtoCaching: Wireless Content Delivery through Distributed Caching Helpers} 
\author{\IEEEauthorblockN{Karthikeyan Shanmugam \IEEEmembership{Student Member,~IEEE} , Negin Golrezaei  \IEEEmembership{Student Member,~IEEE} ,
 Alexandros G. Dimakis \IEEEmembership{Member,~IEEE} }
\IEEEauthorblockN{Andreas F. Molisch  \IEEEmembership{Fellow,~IEEE} , Giuseppe Caire  \IEEEmembership{Fellow,~IEEE} } 
\thanks{Karthikeyan Shanmugam was supported by an Annenberg Fellowship from the University of Southern California during the course of this work.} 
\thanks{Karthikeyan Shanmugam and Alexandros G. Dimakis are with the Department of Electrical and Computer Engineering, University of Texas at Austin, Austin TX 78703 USA.(email: { karthiksh@utexas.edu,dimakis@austin.utexas.edu})}
\thanks{Negin Golrezaei is with the Department of Information and Operations Management, Marshall School of Business, University of Southern California,
Los Angeles 90089 CA USA. (email: { golrezae@usc.edu})}
\thanks{Andreas F. Molisch and Giuseppe Caire are with the Department of Electrical Engineering, University of Southern California,
Los Angeles 90089 CA USA. (email: { molisch@usc.edu,caire@usc.edu}) }
 \thanks{Copyright (c) 2012 IEEE. Personal use of this material is permitted. 
  However, permission to use this material for any other purposes must be obtained from the IEEE by sending a request to pubs-permissions@ieee.org.}
}

\maketitle

\begin{abstract}
Video on-demand streaming from Internet-based servers is becoming one of the most important services offered by wireless networks today. 
In order to improve the area spectral efficiency of video transmission in cellular systems, small cells heterogeneous architectures 
(e.g., femtocells, WiFi off-loading) are being proposed, such that video traffic to nomadic users can be handled 
by short-range links to the nearest small cell access points (referred to as ``helpers''). 
As the helper deployment density increases, the backhaul capacity becomes the system bottleneck. 
In order to alleviate such bottleneck we propose a system where helpers with low-rate backhaul but high storage capacity 
cache popular video files.  Files not available from helpers are transmitted by the cellular base station.  
We analyze the optimum way of assigning files to the helpers,  in order to minimize the expected downloading time for files.  
We distinguish between the uncoded case (where only complete files are stored) and the coded case, where segments 
of Fountain-encoded versions of the video files are stored at helpers. 
We show that the uncoded optimum file assignment is NP-hard, and develop a greedy strategy that is provably 
within a factor 2 of the optimum. Further, for a special case we provide an efficient algorithm achieving 
a provably better approximation ratio of $1-\left(1-1/d \right)^d$, where $d$ is the maximum number of helpers 
a user can be connected to. We also show that the coded optimum cache assignment problem is convex that can be 
further reduced to a linear program. We present numerical results comparing the proposed schemes.\\\\
\end{abstract}


%%%%%%%%%%%%%%%%%%%%%%%%%%%%%%%%%%%%%%%%%%%%%%%%%%%%%%%%
\section{Introduction}
Streaming of video on-demand files is the main reason for the dramatic growth of data traffic over cellular networks -- an increase of two orders of magnitude  compared to current volume is expected in the next five years \cite{cisco66}.  
It is widely acknowledged that conventional current (3G) and near-future (4G-LTE) macro-cell architecture will not be able to support such
traffic increase, even after allocating new cellular spectrum.\footnote{Following a 2010 presidential memorandum in the USA, the government body that regulates wireless frequency bands is releasing  500 MHz of additional bandwidth to wireless networks \cite{Presmemonline}.
This corresponds to approximately doubling the current allocation. 
Since capacity grows linearly with bandwidth, this yields at most a factor of 2 increase in the total system 
capacity while insisting on the current conventional technology.}
In order to tackle such {\em cellular spectrum crunch}, 
the most promising approach to achieve the system area spectral efficiency consists of
shrinking the cell size and essentially {\em bringing the content closer to the users},  
by deploying small base stations that achieve localized communication and enable high-density 
spatial reuse of communication resources \cite{Molisch_2011_book}.
Such pico- and femto-cell networks, which are usually combined with
macrocells into a heterogeneous network, are receiving a lot of attention in the recent literature, (e.g., see \cite{surveyfemto} and references therein). 
A drawback of this approach, though, is the requirement for high-speed backhaul to connect the small cell access points to
the core network \cite{surveyfemto}. In particular, current research trends consider scenarios where the density of small cell
access points will be comparable to the user density \cite{Qualcommwhite1}\cite{Qualcommwhite2} \cite{rusek2013scaling}.
In this regime, deploying high-speed fiber backhaul will be too expensive, and other technologies such as reusing existing copper
(e.g., DSL) or using millimeter-wave  wireless is envisaged. \footnote{This paper was presented in part at \textit{INFOCOM 2012} and \textit{ICC 2012}.}

This paper is motivated by a novel architecture that we have developed in the framework of a wide university-industry 
collaborative project \cite{VAWN1,VAWN2} and articulated in a number of papers covering various aspects of our 
proposal \cite{Infocom,ICC,mingyue-isit13,dilip-isit13,mingyue-jsac,dilip-arxiv}. 
As summarized in \cite{ComMag}, the main idea of this architecture, nicknamed {\em FemtoCaching},  
consists of replacing backhaul capacity with storage capacity at the small cell access points. 
Using caching at the wireless edge, highly predictable bulky traffic (such as video on-demand) can be efficiently handled. 
In this way, the backhaul is used only to refresh the caches at the rate at which the users' demand distribution evolves over time, 
which is a much slower process than the time scale at which the users place their requests for video streaming. 
For example, special nodes with a large storage capacity and only wireless connectivity (no wired backhaul at all) can be deployed 
in a cell and refreshed by the serving cellular base station at off-peak times.
These nodes, henceforth referred to as {\em helpers},  form a wireless distributed caching 
infrastructure.\footnote{Although on-demand video streaming over the wired Internet is 
already handled by {\em Content Delivery Networks} CDNs \cite{passarella2012survey}, we would like to stress the fact that such networks are
formed by caching nodes in the core network, and therefore do not alleviate the backhaul problem to the wireless access points.}

While the overall system design and optimization is a complex problem that would go well beyond the scope of this paper, 
involving research issues at multiple layers, such as the physical layer of a small cell tier architecture
(see \cite{surveyfemto} and references therein)), or the scheduling protocols for video streaming over wireless networks (e.g., see \cite{neely2012wireless, dilip-arxiv,chen2011adaptive}), here we focus on a particular key system aspect:  the content placement 
problem,  \textit{i.e.}, which files should be cached by which helpers, given a certain network topology (helpers-users connectivity) 
and file popularity distribution. In particular, we wish to minimize the expected total file downloading delay (averaged  over the 
users' random demands) over the cache content placement. We consider two variants of the content placement problem. 
In the first case, referred to as the \emph{uncoded content placement problem}, 
only complete files are stored in the helpers caches.   In the second case, we allow {\em intra-session} coding. 
For simplicity, we assume ideal MDS rateless codes,\footnote{An MDS rateless code generates an arbitrarily long sequence of parity symbols
from an information packet of fixed size of $B$ bits,  such that if the decoder obtains any $B$ parity symbols, it can recover the original $B$ information bits.} which can be closely approached in practice by Raptor codes \cite{shokrollahi2006raptor}.  
In this case, the individual identity of the bits that make up a file is not relevant and what matters is how many parity symbols of a given file are retrieved 
from the helpers within the reach of each user.

In both cases, our goal is to optimize the caches allocation assuming that the users' demand distribution (file popularity) and 
the network connectivity graph are known.   
As said before, the file popularity distribution evolves at a time scale much 
slower than the time scale of streaming and it can be learned accurately by monitoring the user activity. Therefore, 
assuming perfect knowledge is justified, at least for the sake of the simplified model treated in this paper. 
The assumption of perfect centralized knowledge of the network topology is more critical, since this changes at the time scale of 
user movements across the network of helper stations, which is typically much denser than a standard cellular network. 
Optimizing the content placement for each given network topology provides therefore an upper bound on the performance of
any adaptive system that tries to track the user mobility. Furthermore, 
in Section \ref{sec:contrib} we show through Monte Carlo simulation that, for a sufficiently dense network with users randomly 
moving across the helper stations, user mobility does not significantly degrade the performance of our system. 
We also offer a non-rigorous intuitive explanation of this behavior.

%%%%%%%%%%%%%%%%%%%%%%%%%%%%%%%%%%%%%%%%%%%%%%%%%%%%%
\subsection{Prior Work}

The idea of using caching to support mobility in networks has been explored in\cite{siris2011,sourlas2010,gaddah2010,vasilakos2012}. 
The main underlying theme behind this body of work is to use caching at wireless access points to enable 
mobility in publish/subscribe networks. In other words, when a user moves from one location to another the delay experienced by the user during ``hand-off'' between two access points can be minimized if the data is cached 
at the access points when the user connects to it. Different procedures for caching have been analyzed with the aim to support mobility.  For more references, we refer the reader to references in\cite{vasilakos2012}. 
In  another line of work, \cite{baev2008,borst2010}, cache placement problems have been considered for the content distribution networks 
which form a distributed caching infrastructure on the wired network. There is a substantial amount of prior work on caching algorithms 
for web and video content (e.g., \cite{cachingrefs2,cachingrefs1,cachingrefs3} and references therein).  These prior works focus on 
wired networks, do not rely on content popularity statistics and do not have the wireless topology aspects that are central in our formulation. 

Coded caching, posed as an index coding problem, has been studied in \cite{maddah2013decentralized,maddah2012fundamental} where a single base station is the only transmitter, the users cache content individually and  
user demands are arbitrary.  Scaling laws of wireless networks with caching was recently explored in \cite{niesen2009caching}. Further, \cite{gitzenis2012asymptotic} focuses on asymptotic scaling laws for joint content replication and delivery in ad-hoc wireless networks with multi-hop relaying, although this paper does not aim at the optimal solution of the cache placement problem, but at obtaining scaling laws for large networks. 

%%%%%%%%%%%%%%%%%%%%%%%%%%%%%%%%%%%%%%%%%%%%%%%%%%%%%%
\subsection{Contributions}\label{sec:contrib}

The contributions of this work are as follows:
\begin{itemize}
\item Uncoded FemtoCaching:  The uncoded distributed caching problem is a special covering problem that involves placing files in helpers to minimize  the total average delay of all users. We show that finding the optimum placement of files is NP-complete. Further,
we express the problem as a maximization of  a submodular function subject to matroid constraints \cite{matroidtheory}, 
and describe approximation algorithms using connections to matroids and covering problems. In particular, we provide a low-complexity greedy algorithm with $1/2$ approximation guarantee. Further, we exhibit another algorithm for a special case, which involves solving a Linear Program 
(LP) with an additional rounding step, that provides $1-(1-1/d)^d$ approximation to the placement problem where $d$ is the maximum number of helpers connected to a user. 
\item Coded FemtoCaching:  The coded distributed caching problem is a convex program and, not surprisingly, 
can be obtained as the convex relaxation of the uncoded problem. Furthermore, it can be reduced to a linear program (LP) by introducing appropriate auxiliary variables. 
\item Numerical results:  We present numerical results regarding the performance of the coded and the uncoded caching schemes under 
idealistic scenarios with topology and link rates inspired by a typical WLAN (WiFi) campus network and a cellular LTE base station 
\cite{Molisch_2011_book}.
%As it has been noted already, we assume that popularity of files changes very slowly compared to the streaming rate. However, the connectivity of %the wireless network evolves at a time scale comparable to that of streaming, for nomadic users moving at walking distance. Hence, adapting the %cache content through low rate backhaul links as the network connectivity changes
%is not practical. To this end we also present numerical results, for the case of limited mobility, where the cache optimization is run at a given current %realization of the network graph, and
%kept constant for a certain time interval, while the users in the network are allowed to move according to a random walk. By comparing the %performance gap between the case where the caches are optimized at each new network graph realization with the suboptimal practical case %where the caches are kept constant for intervals of fixed duration, we numerically evaluate the impact of (limited) mobility on the proposed caching %system.
\end{itemize}

%%%%%%%%%%%%%%%%%%%%%%%%%%%%%%%%%%%%%%%%%%%%%%%%%%%%%%%%%%%%%%%
\section{Distributed caching placement model and assumptions} \label{Caching-prob-defn}

\begin{figure}[ht]
\centerline{\includegraphics[width=4.5cm]{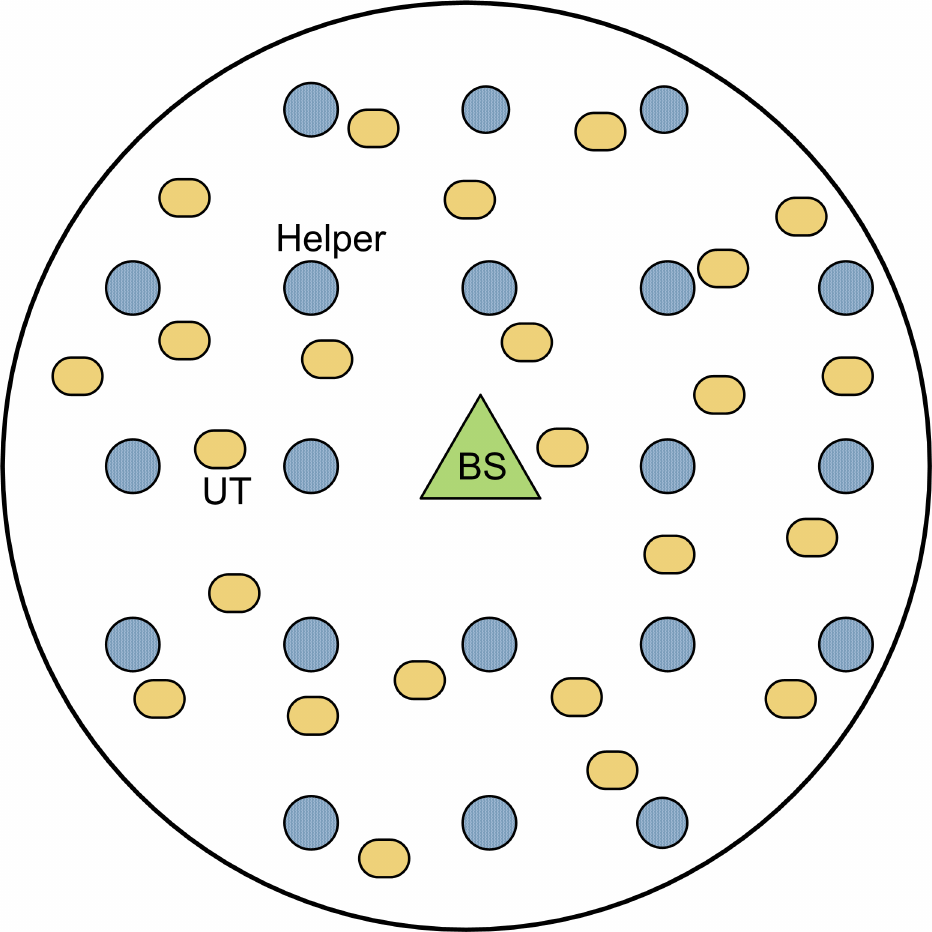}}
\caption{An example of the single-cell layout. UTs are randomly distributed, while helpers can be deterministically placed in the coverage region.}
\label{single-cell1}
\end{figure}

We consider a region where some wireless User Terminals (UTs) place random requests to download files of a 
finite library from a set of dedicated content distribution nodes (helpers). 
The helpers have a limited cache size and limited transmission range, imposing topology constraints both on the 
content placement bipartite graph (involving files and helpers) and on the network connectivity bipartite graph
involving helpers and UTs). 
We also assume the presence of a cellular base station (BS) which contains the whole library 
and can serve all the UTs in the system.  Fig.~\ref{single-cell1} illustrates qualitatively the system layout.  
The key point is that if there is enough content reuse, \textit{i.e.}, many users are requesting 
the same file, caching can replace backhaul communication. 
Notice that requests here are completely asynchronous, i.e., the time difference between 
requests to the same file from different users is arbitrary and generally much larger than the duration of the video playback. 
In contrast, the downloading time should be comparable to the duration of the video playback. Hence, a user can start watching the video
after some (short) time for buffering, while download goes on. 
This assumption is consistent with video on-demand streaming, and prevents direct exploitation of the wireless broadcast medium by  
{\em overhearing} transmissions. In this respect, video on-demand streaming is very different from live streaming, where 
many users wish to get the same content at the {\em same time}. 
 
The content placement problem treated in this work can be formulated as follows: 
{\em for a given file popularity distribution, 
helper storage capacity and network topology, 
how should the files be placed in the helper caches 
such that the average sum downloading delay  of all users is minimized?}

Since users experience shorter delay when they are served locally from helpers in their neighborhoods, 
minimizing the average delay for a given user is equivalent to maximizing the probability of finding the 
desired content in the neighboring helpers. The solution is trivial when there are few helpers in the cell, i.e. when  
each UT can connect only to a single helper. In this case, each helper should cache the most popular files.
However, if the helper deployment is dense enough, UTs will be able to communicate with several such helpers 
and each sees a distributed cache given by the union of the helpers' caches. 
In this situation, the question on how to best assign files to different
helpers becomes a much more complicated and interesting issue, because each UT sees a different, but correlated, distributed cache. 
%In Section \ref{uncoded}, we illustrate through examples the complexity of finding the optimum file placement when some
%users are connected to more than one helper. Moreover, we show that the uncoded distributed caching problem 
%is NP-complete. 

\begin{figure}[ht]	
\centerline{\includegraphics[width=7.5cm]{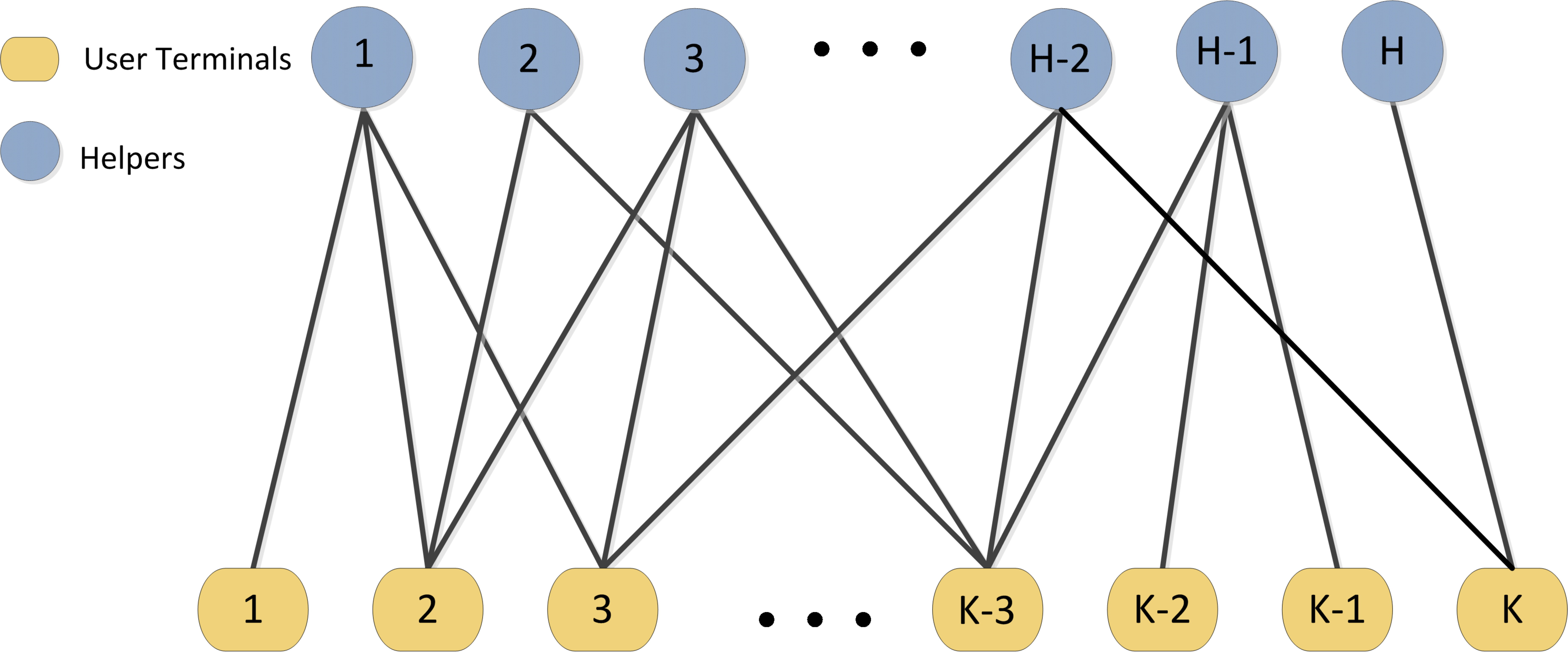}}
\caption{Example of a connectivity bipartite graph indicating how UTs are connected to helpers.}
\label{bipartite}
\end{figure}

We define the set of helpers $\Hc$ of size $H+1$ (where the additional helper is the BS), 
the set of $\Uc$ of size $U$ and a library $\Fc$ comprising $F$ files. 
The wireless network is defined by a bipartite graph $\Gc = (\Hc, \Uc, \Ec)$ (see example in Fig.~\ref{bipartite})
where edges $(h,u) \in \Ec$ denote that a communication link exists from 
helper $h$ to user $u$. We let  $\Uc(h) \subseteq \Uc$ and $\Hc(u) \subseteq \Hc$ denote the sets of 
neighbors of helper $h$ and user $u$, respectively.  
We assume that all users in the system can download from the BS, 
which is conventionally identified with helper $h = 0$. 
Hence, we have $\Uc(0) = \Uc$. The communication links $\{(h,u) : h \in \Hc(u)\}$ are characterized by different rates. 
In particular,  we define the $(H+1) \times U$ matrix $\Omegam$ with elements $\omega_{h,u}$ indicating the {\em average downloading time per information bit} for link $(h,u) \in \Ec$.\footnote{Here, ``average downloading time''
indicates expectation with respect to small-scale effects in the networks, such as frequency and time selective fading, 
and collisions with other transmissions due to some specific MAC layer resource sharing protocol. 
A major simplifying assumption of our model, supported by theoretical results and experimental evidence 
\cite{Borst,shanmugam2012wireless}, is that in current technology wireless networks, under proportional fair resource allocation, 
the average rate of a link depends essentially on the relative position  between transmitter and receiver, 
while the effect of the other nodes in the network produces a ``typical'' impairment 
that does not depend on the specific link, and therefore operates symmetrically on all links.}
 
We assume $\omega_{0,u} \geq \omega_{h,u}$ for all $(h,u) \in \Ec$ (i.e., for all UTs, the delay for downloading from the 
BS is larger than the delay from any other helper), and without loss of generality we set $\omega_{h,u}$ equal to an 
arbitrarily large constant $\omega_{\infty} \gg  \max_u \omega_{0,u}$,  for all $(h,u) \notin \Ec$ (i.e., non-existing links can be 
regarded as links for which downloading one bit of information takes an arbitrarily large amount of time). 
Without loss of fundamental generality, we assume that the files in the library $\Fc$  have the same size of $B$ bits. 
This assumption is mainly used for notational convenience, and could be easily lifted by considering a finer packetization, 
and breaking longer files into blocks of the same length. A probability mass function $\{P_f : f = 1,\ldots, F\}$ is defined on $\Fc$, and 
we assume that users make independent requests to the files $f \in \Fc$ with probability $P_f$.  

%%%%%%%%%%%%%%%%%%%%%%%%%%%%%%%%%%%%%%%%%%%%%%%%%%%%%%%%%%%%%%
\section{Uncoded content placement} \label{uncoded}

%In the uncoded case, a file can be either entirely cached or not cached at all. 
%As mentioned in Section \ref{Caching-prob-defn}, 
%if each UT can communicate to only one helper, the optimal caching policy is simple: each helper should cache the most popular files. 
%When a user has connection to multiple helpers, however, the caching policy becomes non-trivial, as shown in the example of Figure 
%\ref{fg_DistCaching}. 
%There are two helpers and four UTs. The dashed circles  centered around helpers indicate the helper transmission radius.  
%Assuming that each helper can cache $M$ files, users $u_1$ and $u_2$ would prefer helper $h_1$ to cache the $M$ most popular files since 
%this minimizes their average delay.  Similarly, user $u_4$ would prefer that helper $h_2$ also caches the $M$ most popular files. 
%However $u_3$ would prefer $h_1$ to cache the $M$ most popular files and $h_2$ the second $M$ most popular (or the opposite). 
%This effectively creates a distributed cache of size $2M$ for user $u_3$.  We observe that in the distributed caching problem 
%the individual objectives of different users may be in conflict. 
%
%\begin{figure}[ht]	
%\centerline{\includegraphics[width=7cm]{distributed_caching_exam1.pdf}}
%\caption{Distributed Caching example: two helpers and four users with conflicting interests. }
%\label{fg_DistCaching}
%\end{figure}

An uncoded cache placement is represented by a bipartite graph $\widetilde{\Gc} = (\Fc, \Hc, \widetilde{\Ec})$ such that 
an edge $(f,h) \in \widetilde{\Ec}$ indicates that a copy of file $f$ is contained in the cache of helper $h$. 
We let $\Xm$ denote the $F \times H$ adjacency matrix of $\widetilde{\Gc}$, such that $x_{f,h} = 1$ if $(f,h) \in \widetilde{\Ec}$ and 0 otherwise. 
By the cache size constraint, we have that the column weight of $\Xm$ is at most $M$. 

Consider a user $u$ and its helper neighborhood $\Hc(u)$. 
We sort the link delays $\omega_{h,u}$ in increasing order such that $(j)_u$ denotes the helper index with the $j$-th smallest delay
to user $u$. By assumption, we have $(|\Hc(u)|)_u = 0$ (the BS has the highest delay among all $h \in \Hc(u)$ and therefore it is sorted in 
$|\Hc(u)|$-th position by the {\em helper sorting function} $(\cdot)_u$) and for all $j > |\Hc(u)|$ we have $\omega_{(j)_u,u} = \omega_\infty$.~\footnote{By construction, 
the links $((j)_u,u)$ with $j > |\Hc(u)|$ do not exist in $\Ec$.}
With this notation, the average delay per information bit for user $u$ can be written as:~\footnote{We use the convention that
the result of $\prod_{i=a}^b$ is 1 when $b < a$, and that the result of $\sum_{i=a}^b$ is zero when $b < a$.}
\begin{align}
\bar{D}_u  = & \;  \sum_{j=1}^{|\Hc(u)|-1} \omega_{(j)_u,u} \sum_{f=1}^F \left [ \prod_{i=1}^{j-1} (1 - x_{f,(i)_u}) \right ] x_{f, (j)_u} P_f 
  \nonumber \\
 \hfill   &+ \omega_{0,u} \sum_{f=1}^F \left [ \prod_{i=1}^{|\Hc(u)|-1} (1 - x_{f,(i)_u}) \right ] P_f.  \label{Dk}
\end{align}
In order to see this, notice that $ \left [ \prod_{i=1}^{j-1} (1 - x_{f,(i)_u}) \right ] x_{f, (j)_u}$ is the indicator function (defined over the set of feasible placement matrices $\Xm$)
for the condition that file $f$ is in the cache of the helper $(j)_u$ (the $j$-th lowest delay helper for user $u$), and it is not in any of the helpers 
with lower delay $(i)_u$, for $i = 1,\ldots, j-1$. Also, $\left [ \prod_{i=1}^{|\Hc(u)|-1} (1 - x_{f,(i)_u}) \right ]$ is the indicator function
for the condition that file $f$ is not found in the neighborhood $\Hc(u) \backslash \{0\}$ of user $u$. 

The minimization of the sum (over the users) average per-bit downloading delay can be expressed as the following integer programming problem: 
\begin{eqnarray} \label{max problem}
\mbox{maximize} & & \sum_{u=1}^{U} \left ( \omega_{0,u} - \bar{D}_u \right ) \nonumber \\
\mbox{subject to} & & \sum_{f=1}^F x_{f,h} \leq M, \;\;\; \forall \;\; h,  \nonumber \\
& & \Xm \in \{0,1\}^{F \times H}. 
\end{eqnarray}
In the following, we will show that problem (\ref{max problem}) is NP-complete. 
Then, we  formulate the problem as maximization of a monotone submodular function over matroid constraints. 
This structure allows us to use a greedy strategy for placement which provably achieves at least $\frac{1}{2}$ 
of the optimum value, uniformly over
all instances of the problem.  Moreover, for a special class of this problem, we present an algorithm that gives even a better performance guarantee.

%%%%%%%%%%%%%%%%%%%%%%%%%%%%%%%%%%%%%%%%%%%%%%%%%%%%%%%%%%%%%%%% 
\subsection{Computational intractability}

To show that the optimization problem in (\ref{max problem}) is NP-complete, we consider a special case of the problem 
where $\omega_{h,u} = \omega_1 \geq 0$ for all $(h,u) \in \Ec$ with $h \neq 0$, and $\omega_{0,u} > \omega_1$ for all $u$. 
In this case, letting $\widetilde{\omega}_u = \omega_{0,u} - \omega_1$, (\ref{max problem}) becomes:
\begin{eqnarray} 
\mbox{maximize} & & \sum_{f=1}^F P_f \sum_{u=1}^{U} \widetilde{\omega}_u  \left [ 1  -  \prod_{h\in \Hc(u):h\neq 0} (1 - x_{f,h}) \right ] \nonumber \\
\mbox{subject to} & & \sum_{f=1}^F x_{f,h} \leq M, \;\;\; \forall \;\; h,  \nonumber \\
& & \Xm \in \{0,1\}^{F \times H}  \label{eqn:IPform2}
\end{eqnarray}
%
%\begin{eqnarray} 
%\mbox{maximize} & & \sum_{f=1}^F \sum_{u=1}^{U}  \sum_{f  \in \Ac_u} P_f  \nonumber \\
%\mbox{subject to} & & \sum_{f=1}^F x_{f,h} \leq M, \;\;\; \forall \;\; h,  \nonumber \\
%& & \Xm \in \{0,1\}^{F \times H} \nonumber \\ 
%& & \Ac_u = \left \{ f : \sum_{h \in \Hc(u) : h \neq 0} x_{f,h} > 0 \right \}. \label{Ak}
% \end{eqnarray}
where $\left [ 1  -  \prod_{h\in \Hc(u):h\neq 0} (1 - x_{f,h}) \right ]$ is the indicator function (over the set of feasible $\Xm$) 
of the condition $f \in \Ac_u$, and $\Ac_u$  is the union of the helpers' caches in the neighborhood of user $u$, excluding the BS. 
The above objective function can be interpreted as the sum of \emph{values} seen by each user. 
The value of each user $u$ is equal to $\widetilde{\omega}_u \sum_{f \in \Ac_u} P_f$, which is proportional to the probability of 
finding a file in the union of the helpers' caches, multiplied by the incremental delay to download such files from the BS rather than 
from the helpers.  Our goal here is to maximize the sum of values seen by all users.  
To prove that the problem is NP-hard, we consider its corresponding 
decision problem, referred to as the {\em Helper Decision Problem}.

\begin{problem}\label{Prob:helper}
   (\textit{Helper Decision Problem})
Given the network connectivity graph $\Gc = (\Uc, \Hc, \Ec)$, the library of files $\Fc$, 
the popularity distribution $\Pc = \{P_f\}$, the set of positive real numbers
$\Omega = \{ \widetilde{\omega}_u : u \in \Uc\}$ and a real number $Q \geq 0$, 
determine if there exists a feasible cache placement $\Xm$ with cache size constraint $M$ such that
\begin{equation}\label{eqn:utility-fn}
\sum_{u=1}^{U} \widetilde{\omega}_u \sum_{f  \in \Ac_u} P_f   \geq Q.
\end{equation}
Let the problem instance be denoted by ${\rm HLP}(\Gc,\Fc,\Pc,\Omega,M,Q)$.  \hfill $\lozenge$
\end{problem}

It is easy to see that the helper decision problem is in the class NP. 
To show NP-hardness, we will use a reduction from the following NP-complete problem.

\begin{problem} \label{Prob:2DSC}
(\textit{2-Disjoint Set Cover Problem}) 
Consider a bipartite graph $G = (A,B,E)$ with edges $E$ between two disjoint vertex 
sets $A$ and $B$. For $b \in B$,
define the neighborhood of $b$ as $\Nc(b) \subseteq A$. Clearly, $A = \bigcup
\limits_{b \in B} \Nc(b)$. Do there exist two disjoint sets $B_{1},B_{2} \subset B$ such that 
$|B_{1}|+|B_{2}|=|B|$ and $A = \bigcup
\limits_{b \in B_{1} } \Nc(b)  = \bigcup
\limits_{b \in  B_{2} } \Nc(b)$? 
Let the problem instance be denoted by ${\rm 2DSC}(G)$. \hfill $\lozenge$
\end{problem}

It is known that the 2-disjoint set cover problem is NP-complete~\cite{Cardei}. 
We show in the following lemma that given a unit time oracle for the helper decision problem, 
we can solve the 2-disjoint set cover problem in polynomial time (a polynomial time reduction is denoted by $\leq_{L}$).

\begin{lemma}
\textit{2-Disjoint Set Cover Problem} $\leq_L$ \textit{Helper Decision Problem}.
\end{lemma}

\begin{figure}[ht]
 \centering
 \includegraphics[scale=0.45]{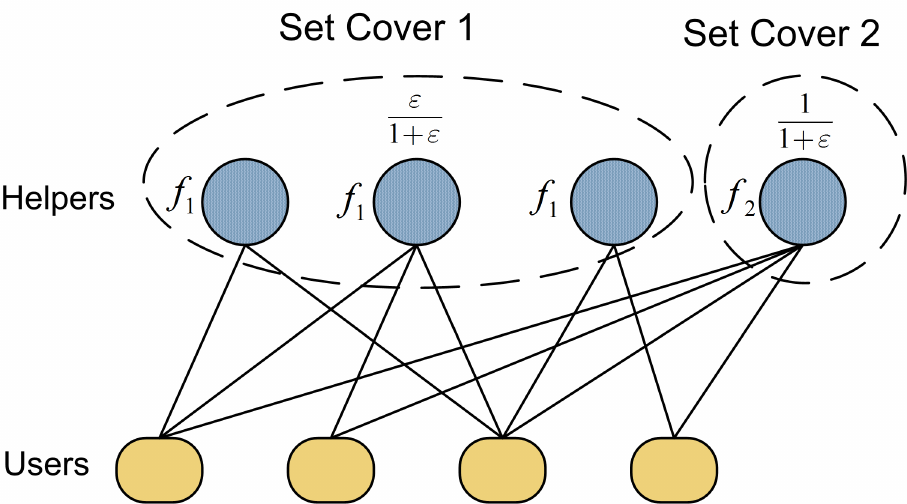}
  \caption{Figure illustrating the reduction from 2-Disjoint Set Cover Problem.}
 \label{fig:Helper}
\end{figure}

\begin{IEEEproof}
Consider an oracle that can solve any instance ${\rm HLP}(\Gc,\Fc,P,\Omega,M,Q)$ in unit time. 
Then solving an instance of ${\rm 2DSC}(G)$ is equivalent to solving 
${\rm HLP}(\Gc,\Fc, \Pc,\Omega,M,Q)$ for
$\Gc = G$, $\Fc = \{1,2\}$, 
$\Pc = \left (\frac{1}{1+\epsilon}, \frac{\epsilon}{1+\epsilon}\right )$,
$\Omega = \{1,1,\ldots, 1\}$, $M = 1$ and $Q = U = |\Uc|$. 

In order to see this, notice that for any user $u$ we have that $\Ac_u$ can be $\emptyset, \{1\}, \{2\}, \{1,2\}$. 
Notice also that the value of any user $u$ is $\sum_{f \in \Ac_u} P_f \leq 1$ and it is exactly equal to 1 only if
$\Ac_u = \{1,2\}$. However, since the cache constraint for all helpers is equal to 1, any helper can cache either file 1 or file 2 or none. 
It follows that in order to have the objective function value equal to $U$ the value of all users must be equal to 1, i.e., 
$\Ac_u = \{1,2\}$ for all $u$, which implies that each user $u$ has at least one neighboring helper containing file 1 and another neighboring helper
containing file 2. Letting $B_1$ and $B_2$ denote the (disjoint) sets of helpers containing file 1 and file 2, respectively, 
we conclude that determining whether the objective function (\ref{eqn:utility-fn}) is equal to $U$ is equivalent to determining 
the existence of $B_1$ and $B_2$ forming a 2-disjoint cover (see Fig.~\ref{fig:Helper}). 
\end{IEEEproof}
 
%%%%%%%%%%%%%%%%%%%%%%%%%%%%%%%%%%%%%%%%%%%%%%%%%%%%%%
\subsection{Computationally efficient approximations}

In this section, we show that Problem \ref{max problem} can be formulated as the maximization of a submodular function subject to
matroid constraints. This structure can be exploited to devise computationally efficient algorithms for 
Problem \ref{max problem} with provable approximation gaps.
The definitions of matroids and submodular functions are reviewed in Appendix \ref{app:matroids}.
First, we define the following ground set:
\begin{equation}\label{ground set}
S = \{s_{1}^{1},s_{2}^{1}, \ldots , s_{F}^{1}, \ldots , s_{1}^{H},s_{2}^{H}, \ldots ,s_{F}^{H} \}, 
\end{equation}
where $s_f^h$ is an abstract element denoting the placement of file $f$ into the cache of helper $h$. 
The ground set can be partitioned into $H$ disjoint subsets, 
$S_{1} , \ldots , S_{H}$, where $S_{h} = \{s_{1}^{h},s_{2}^{h}, \ldots , s_{F}^{h}\}$ is the set of all files that might be placed in the cache 
of helper $h$.

\begin{lemma}
The constraints in (\ref{max problem}) can be written as a partition matroid on the ground set $S$ defined in (\ref{ground set}).
\end{lemma}

\begin{IEEEproof}
In (\ref{max problem}), a cache placement is expressed by the adjacency matrix $\Xm$. 
We define the corresponding cache placement set $X \subseteq S$  such that 
$s_f^h \in X$ if and only if $x_{f,h} = 1$. Notice that the non-zero elements of the $h$-th column of $\Xm$ correspond to the elements in $X \cap S_h$. 
Hence, the constraints on the cache capacity of helpers can be expressed as $X \subseteq \Ic$, where
 \begin{equation}\label{our matroid}
 \Ic = \{ X \subseteq S : |X \cap {S_h}| \le M,\;\;\forall \; h = 1, \ldots ,H \}.
 \end{equation} 
Comparing $\Ic$ in (\ref{our matroid}) and the definition of the partition matroid in (\ref{partition}), 
we can see that our constraints form a partition matroid with $l = H$ and $k_{i} = M$, for $i=1,...,H$. 
The partition matroid is denoted by $\Mc = (S, \Ic)$.
\end{IEEEproof}

Notice that the set $\{x_{f,h} : f \in \Fc\}$ can be considered as the Boolean representation of $X_h = X \cap S_h$, in the sense that
$x_{f,h}  = 1$ if $s_f^h \in X_h$ and $x_{f,h} = 0$ otherwise.  We have:

\begin{lemma}
The objective function in Problem (\ref{max problem}) is a monotone submodular function.
\end{lemma}

\begin{IEEEproof}
Monotonicity is obvious since any new placement of a file cannot decrease the value of the objective function.
In order to show submodularity, we observe that since the sum of submodular functions is submodular, it is enough to prove that for a user $u$ the set function  $G_u(X) \triangleq \omega_{0,u} - \bar{D}_u$ is submodular.
We show that the marginal value of adding a new file to an arbitrary helper $h \in \Hc(u)$ decreases as the placement set $X$ 
becomes larger. The marginal value of adding a new element to a placement set $X$ is the amount of increase 
in $G_u(X)$ due to the addition.

Let's consider two placement sets $X$ and $X'$  where $X \subset X' \subset S$.  For some $1 \leq i \leq |\Hc(u)| - 1$, 
consider adding the element $s_f^{(i)_u} \in S \backslash X'$ to both placement sets. 
This corresponds to adding file $f$ in the cache of helper $(i)_u$, where
such file is not placed anywhere neither in placement $X$ nor in placement $X'$. 
We distinguish the following cases. 

1) According to placement $X'$, user $u$ gets file $f$ from helper $(j')_u$ with $j' < i$, i.e., 
$s_f^{(j')_u} \in X'$.  In this case, it is immediate to see that $G_u(X' \cup \{s_f^{(i)_u}\}) - G_u(X') = 0$ (the marginal value is zero). 
According to placement $X$, user $u$ gets file $f$ from some helper $(j)_u$  with  $j \geq j'$. 
If $j < i$, again the marginal value will be zero. However, when $j > i$, the marginal value is given by 
$G_u(X \cup \{s_f^{(i)_u}\}) - G_u(X) = P_f (\omega_{(j)_u,u} - \omega_{(i)_u,u}) > 0$.

2) According to placement $X'$, user $u$  gets file $f$ through helper $(j')_u$ with $j' > i$. Hence, the marginal value is given by 
$G_u(X' \cup \{s_f^{(i)_u}\}) - G_u(X') = P_f (\omega_{(j')_u,u} - \omega_{(i)_u,u})$.
Since in the placement set $X$, user $u$ downloads the file from helper $(j)_u$ with $j \geq j'$, the resulting marginal value
is $G_u(X \cup \{s_f^{(i)_u}\}) - G_u(X) = P_f (\omega_{(j)_u,u} - \omega_{(i)_u,u})$. The difference of marginal values is given by: 
\begin{align*}
 \hfill &G_u(X \cup \{s_f^{(i)_u}\}) - G_u(X) - \left ( G_u(X' \cup \{s_f^{(i)_u}\}) \right. \\
 \hfill & \left. - G_u(X') \right )=  P_f ( \omega_{(j)_u,u}  - \omega_{(j')_u,u} ) \geq 0. 
\end{align*}
Hence, the lemma is proved.
\end{IEEEproof}

A common way to maximize a monotonically non-decreasing submodular function subject to a matroid constraint  consists of a greedy algorithm
that starts with an empty set and at each step it adds one element with the highest marginal value to the set while maintaining the feasibility of the solution. 
Since the objective function is submodular, the marginal value of elements decreases as we add more elements to the placement set $X$. 
Thus, if at one iteration, the largest marginal value is zero, then the algorithm should stop.  For our case, there would be $MH$ iterations till all the caches are filled.
Each iteration would involve, evaluating marginal value of at most $FH$ elements. Each evaluation takes $O(U)$ time. Hence the running time would be $O(F^2H^2 U)$.
Classical results on approximation of such maximization problems \cite{greedy} establish 
that the greedy algorithm achieves an objective function value within a factor $\frac{1}{2}$ of the optimum.

For maximization of a general monotone submodular function subject to matroid constraints, 
a randomized algorithm which gives a $(1 -1/e)$-approximation has been proposed in \cite{multilinear}. 
This algorithm consists of two parts. In the first part, the combinatorial integer programming 
problem is replaced with a continuous one and an approximate  solution of the continuous problem is found. 
In the second part, the found feasible point  of the continuous problem is rounded using a technique called pipage rounding~\cite{multilinear}

Although this algorithm gives a better performance guarantee than greedy placement, 
when $|S| = H F$ becomes large, its complexity is still too computationally demanding for implementation. 
Specifically, the running time of the algorithm in \cite{multilinear} is $O(n^8)$ where $n$ is rank of the matroid. 
In our formulation, the rank of the matroid is $MH$. 
Hence, the time complexity is $O(MH)^8$.  When $M$ is a constant fraction of $F$, the time complexity is $O((HF)^8)$.
In practice, it is reasonable to assume $M \sim 100$ and $F \sim 1000$ (e.g., the 1000 most popular titles of the Netflix library), while
$H$ may range from 10 to 100, for a 1 sq.km cell. 
Hence, running this algorithm is not just impractical for a real-time system implementation, but even for a computer simulation on a powerful server. 
As a matter of fact, for all practical purposes, the greedy algorithm that maximizes at each step the marginal value
is the most suitable for the most general case of the problem, because of the typically large problem size. 
However, in the special case when $\omega_{h,u} = \omega_1$ (fixed value independent of $(h,u)$) for all $(h,u) \in \Ec, h \neq 0$, 
we provide a different approximation algorithm that runs in $O\left( (U + H)^{3.5} F^{3.5}\right)$ time and provides an approximation ratio of $1-\left(1-1/d \right)^d$, 
where $d = \max_u\{ |\Hc(u)| - 1\}$ is the maximum number of helpers a user is connected to in $\Gc$ (excluding the BS). 
When no bounds on $d$ can be established, this algorithm recovers the ratio of $1-1/e$. 
We mention that although the worst case time complexity guarantee is $O( (U + H)^{3.5} F^{3.5})$ ,
the algorithm  involves solving a linear program (LP) with $O((U+H)F)$ variables and a simple deterministic rounding algorithm. The worst cast time complexity of the algorithm
is dominated by the time complexity of interior point methods that could be used for the LP involved. But it is well-known that for most cases, LP runs much faster than the time complexity bound offered by the interior-point methods. 

%%%%%%%%%%%%%%%%%%%%%%%%%%%%%%%%%%%%%%%%%%%%%%%%
%%%%%%%%%%%%%%%%%%%%%%%%%%%%%%%%%%%%%%%%%%%%%%%%
%%%%%%%%%%%%%%%%%%%%%%%%%%%%%%%%%%%%%%%%%%%%%%%%
%%%%%%%%%%%%%%%%%%%%%%%%%%%%%%%%%%%%%%%%%%%%%%%%
\subsection{Improved Approximation ratio for the uncoded problem}\label{sec:specialcase}

In this section, we provide an improved approximation algorithm for the uncoded caching problem in the special case where 
$\omega_{h,u} = \omega_1$ for all $(h,u) \in \Ec$ with $h \neq 0$, and $\omega_1 < \omega_{0,u}$ for all $u \in \Uc$. Recall that, in this case, 
the optimization problem is given by (\ref{eqn:IPform2}). For convenience, 
%Here, we rewrite this problem in the following equivalent form:
%\begin{eqnarray} 
%\mbox{maximize} & & \sum_{f=1}^F P_f \sum_{u=1}^{U} (\omega_{0,u} - \omega_1)  \left [ 1  -  \prod_{h\in \Hc(u):h\neq 0} (1 - x_{f,h}) \right ] \nonumber \\
%\mbox{subject to} & & \sum_{f=1}^F x_{f,h} \leq M, \;\;\; \forall \;\; h,  \nonumber \\
%& & \Xm \in \{0,1\}^{F \times H}  \label{eqn:IPform2}
%\end{eqnarray}
we define $g_{f,u}(\Xm) =  \left [ 1  -  \prod_{h\in \Hc(u):h\neq 0} (1 - x_{f,h}) \right ]$ and write the objective function in
(\ref{eqn:IPform2}) as
\begin{equation}\label{ggg}
g(\Xm) =  \sum_{u,f} \; P_f \; \widetilde{\omega}_u \; g_{f,u}(\Xm),
\end{equation} 
The program (\ref{eqn:IPform2}) fits the general framework of maximizing a function subject 
to integral assignment constraints involving assignment variables ($\Xm$ in our case) 
corresponding to the edges of a bi-partite graph, ($\widetilde{\Gc}$ in our case). 
This general framework has been studied in \cite{ageev2004pipage}.  
The authors of \cite{ageev2004pipage} provide sufficient conditions under which the optimum of a relaxation of a suitable problem \textit{related} to  
(\ref{eqn:IPform2}) can be rounded using the technique of \textit{pipage rounding} to achieve a constant 
approximation guarantee for problem (\ref{eqn:IPform2})  .  In \cite{ageev2004pipage}, this was carried out for the maximum coverage problem (or max $k$-cover problem), i.e., 
the problem of choosing $k$ sets out of a fixed collection of $m$ subsets of some ground set in order to maximize the number of covered elements of the ground set.
Our uncoded caching problem, in the special case considered in this section, is similar in structure to the maximum coverage problem, 
although we are not aware of any reduction between the two problems. 
However, this structural similarity allows us to apply the tools developed in \cite{ageev2004pipage} in order to obtain a constant 
factor approximation to problem (\ref{eqn:IPform2}). 
The result heavily hinges on the machinery developed in \cite{ageev2004pipage}.  
We state results from \cite{ageev2004pipage} and outline the relevant proofs 
for the sake of clarity.

We first describe the pipage rounding technique for the following template problem on the bi-partite graph 
$G = (A,B,E)$, where the matrix $\Rm \in \RR_+^{|A| \times |B|}$ contains the optimization variables $\rho_{a,b}$ for all edges $(a,b) \in E$ (these variables are
fixed to 0 for the elements $(a,b) \notin E$).  
\begin{eqnarray} 
\mbox{maximize} & & \phi(\Rm)  \label{eqn:tempprog} \\
\mbox{subject to} & & \sum_{a : (a,b) \in E} \rho_{a,b} \leq p(b), \;\;\; \forall \;\; b,  \label{eqn:tempcon1}  \\
\mbox{subject to} & & \sum_{b: (a,b) \in E} \rho_{a,b} \leq p(a), \;\;\; \forall \;\; a,   \label{eqn:tempcon11}  \\
& & \Rm \in [0,1]^{|A| \times |B|}  \label{eqn:tempcon2} 
\end{eqnarray}

Here, $p(y) \in \mathbb{Z}_{+}, ~ \forall y$. Observe that it is the relaxed version of an integer program where the objective function is $\phi(\mathbf{X})$,  variables $\rho_{a,b}$ are replaced by $x_{a,b} \in \{ 0,1\}$ and $\mathbf{X} \in \{0,1\}^{|A| \times |B|}$ . The pipage rounding algorithm takes as input $\phi,G,\{p(y)\}$ and a real feasible solution $\mathbf{R}$ and outputs a feasible integral solution $\mathbf{\bar{X}}$.  For the sake of completeness and subsequent use in the arguments later, this procedure is provided as Algorithm \ref{alg:pipageround} in the paper.
     
\begin{algorithm}
\caption{PipageRounding$(G,\phi,\{p(y)\},\mathbf{R})$} 
\label{alg:pipageround}
\begin{algorithmic}
	\STATE Input:  $\mathbf{R}$. Output: $\mathbf{\bar{X}}$.
	\STATE Initialize: $\mathbf{\hat{X}}= \mathbf{R}$ \\
	\WHILE{$\mathbf{\hat{X}} \mathrm{~is~ not~ integral}$ }
		\STATE Construct the subgraph $H_{\hat{x}}$ of $G$ which contains all vertices but the edge set $E_{\hat{x}}$ is such that $(a,b) \in E_{\hat{x}}$ only if $\hat{x}_{a,b}$ is not integral  \\
		\IF {$H_{\hat{x}}$ has cycles}
		       \STATE Set $\alpha$ to be one such simple cycle (each node occurs at most once in a simple cycle except the first node)\\
		 \ELSE
		        \STATE Set $\alpha$ to be a simple path with end points being nodes of degree 1 (In a  bi-partite cycle-free graph, such a path exists).\\
		  \ENDIF           
		\STATE Decompose $\alpha = M_1 \bigcup M_2$ (union of two disjoint matchings) where $M_1$ and $M_2$ are matchings.\\
	          \STATE Define parameterized solution $\mathbf{X}(\epsilon,\alpha)$ as follows: \\
		\STATE If $(a,b) \notin \alpha$, $x_{a,b}(\epsilon,\alpha) = \hat{x}_{a,b}$. \\
		\STATE Let $\epsilon_1=\min \{ \min \limits_{(a,b) \in M_1} \hat{x}_{a,b}, \min \limits_{(a,b) \in M_2} (1-\hat{x}_{a,b}) \}$. \\
		\STATE Let $\epsilon_2= \min \{\min \limits_{(a,b) \in M_2} \hat{x}_{a,b}, \min \limits_{(a,b) \in M_1} (1-\hat{x}_{a,b}) \}$.\\
		\STATE $x_{a,b}(\epsilon,\alpha)= \hat{x}_{a,b}+\epsilon, ~\forall (a,b) \in M_1$ and $x_{a,b}(\epsilon,\alpha)= \hat{x}_{a,b}-\epsilon, ~\forall (a,b) \in M_2$. $\epsilon \in \left[-\epsilon_1,\epsilon_2 \right]$. \\
		\IF {$\phi \left(\mathbf{X}\left(\epsilon_2,\alpha \right) \right) > \phi \left(\mathbf{X}\left(-\epsilon_1,\alpha \right)\right) $} 
		     \STATE $\mathbf{\hat{X}} = \mathbf{X} \left(\epsilon_2,\alpha \right)$  \\
		\ELSE
		    \STATE $\mathbf{\hat{X}}=  \mathbf{X} \left(-\epsilon_1,\alpha \right) $   \\        
		\ENDIF    
	\ENDWHILE  
	\STATE return $\mathbf{\bar{X}}=\mathbf{\hat{X}}$.
\end{algorithmic}
\end{algorithm}        
    
We recall the definition of a \textit{matching} used in Algorithm \ref{alg:pipageround}. A matching in an undirected graph is a subset of edges such that no two edges in the subset have a common vertex.  Algorithm \ref{alg:pipageround} runs in at most $\lvert E \rvert$ steps (a step being the outer while loop). Each step runs in time polynomial in $\lvert E \rvert$. The final solution $\mathbf{\bar{X}}$ is integral and it satisfies the constraints in program (\ref{eqn:tempprog})--(\ref{eqn:tempcon2}). We refer the reader to \cite{ageev2004pipage} for the proof of correctness and the time complexity analysis for pipage rounding.  Now, we have following sufficient conditions for a $C-$ approximate polynomial time algorithm that uses pipage rounding.

\begin{theorem}\label{Thm:FL} \cite{ageev2004pipage}
Consider the problem (\ref{eqn:tempprog}) -- (\ref{eqn:tempcon2}) and suppose that $\phi(\cdot)$ satisfies the following conditions:
\begin{enumerate}
            \item There exists another objective function $L(\cdot)$ such that $ \forall ~\Xm \in \{0,1\}^{|A| \times |B|}$,  $L(\mathbf{X})=\phi(\Xm)$.
            \item (\textit{Lower Bound condition}) For $\Rm \in \RR_+^{|A| \times |B|}$, $\phi \left( \Rm \right) \geq C L \left( \Rm \right) $ for some constant $C < 1$.
             \item(\textit{$\epsilon$-convexity condition}) For all feasible $\Rm$ in (\ref{eqn:tempprog}) -- (\ref{eqn:tempcon2}) 
and for all possible cycles and paths $\alpha$ that occur in pipage rounding , $\phi \left(\mathbf{X}\left(\epsilon,\alpha \right) \right) $ is convex with respect to $\epsilon$ in the range $\epsilon \in \left[ -\epsilon_1, \epsilon_2 \right]$, where $\mathbf{X}\left(\epsilon,\alpha \right),\epsilon_1,\epsilon_2$ are intermediate values that occur in every iteration as defined in Algorithm \ref{alg:pipageround}.
        \end{enumerate}
Let the optimum of the maximization $\max \;  L(\Rm)$ subject to  (\ref{eqn:tempcon1}), (\ref{eqn:tempcon11}), (\ref{eqn:tempcon2}), 
be $\Rm_{opt}$. Let $\Xm_{int}$ be the integral output of ${\rm PipageRounding} \left(G,\phi,\{p(u) : u \in \Ac \cup \Bc \} , \Rm_{opt} \right)$ (see Algorithm \ref{alg:pipageround}).
Then, $\phi \left(\Xm_{int} \right) \geq C \phi \left( \Xm_{\rm opt} \right)$ where $\Xm_{\rm opt}$ 
is the optimum solution to the integer version of program (\ref{eqn:tempprog}) -- (\ref{eqn:tempcon2}), obtained by 
replacing $\Rm$ with the binary matrix $\Xm$.       
 \end{theorem} 
     
\begin{IEEEproof}
      At the end of the inner while loop of Algorithm \ref{alg:pipageround}, one of the end points of the curve $\mathbf{X} (\epsilon,\alpha)$ is chosen as the improved solution. The next iteration proceeds with this improved solution. For $\epsilon=0 \in \left[ -\epsilon_1, \epsilon_2 \right]$, $\mathbf{X}(0,\alpha) = \mathbf{\hat{X}}$ where $\mathbf{\hat{X}}$ is the solution from the previous iteration. If $\phi \left( \mathbf{X}(\epsilon,\alpha) \right)$ is convex in $\epsilon$ ($\epsilon$- convexity condition), then the maximum is attained at the end points. Therefore,  $\max \{ \phi \left(\mathbf{X} (-\epsilon_1,\alpha)\right), \phi \left(\mathbf{X} (\epsilon_2,\alpha)\right) \} \geq \phi \left( \mathbf{\hat{X}} \right)$. Hence, the solution at the end of the inner while loop is no worse than the solution at the beginning. Therefore, if $\mathbf{R}_{opt}$ is the input to the pipage algorithm and $\mathbf{X}_{int}$ is the output, then we have the following chain of inequalities:         
      \begin{align*}     
         \phi \left( \mathbf{X}_{int} \right) & \overset{(a)}{\geq} \phi \left( \mathbf{R}_{opt} \right) \\
         \hfill &  \overset{(b)}{\geq} C L \left(\mathbf{R}_{opt} \right) \overset{(c)}{\geq} C L \left(\mathbf{X}_{opt} \right)  \overset{(d)}{\geq} C \phi \left( \mathbf{X}_{opt} \right) 
       \end{align*}
       Justification for the above inequalities are: (a) pipage rounding with $\epsilon$-convexity condition ;(b) Lower Bound condition ;(c) $\mathbf{R}_{opt}$ is obtained when $L(\cdot)$ is optimized over the reals (relaxed version of program (\ref{eqn:tempprog})--(\ref{eqn:tempcon2}) with $L$ as the objective) ;(d)  Condition $1$ in the theorem.                
     \end{IEEEproof}

In our case, it is easy to particularize the general template program (\ref{eqn:tempprog}) -- (\ref{eqn:tempcon2}) to
the program at hand (\ref{eqn:IPform2}), by letting  $\phi(\cdot) = g(\cdot)$, defined in (\ref{ggg}), 
by identifying the graph $G$ with  the complete bipartite graph ${\cal K}_{\Fc,\Hc}$ formed by the vertices $\Fc$, $\Hc$ and all possible edges connecting 
the elements of $\Fc$ (files) with the elements  of $\Hc$ (helpers), and the edge node constraints as $p(h) = M$ for all $h \in \Hc$ and $p(f) = H$ for all $f \in \Fc$. 
Notice that, any feasible placement graph $\widetilde{\Gc}$ is a subgraph of this complete bipartite graph, and that letting
$p(f) = H$ makes the set of constraints (\ref{eqn:tempcon11}) irrelevant.

Now, we design a suitable $L(.)$ function such that $\phi \left( \mathbf{X} \right) = L \left( \mathbf{X} \right)$ $\forall ~ \Xm \in \{ 0,1\}^{|\Fc| \times |\Hc|}$. Let $L=\sum \limits_{f,u} P_f   \widetilde{\omega}_u L_{f,u} \left( \Xm \right)$ and $L_{f,u} \left(\mathbf{X} \right) = \min \{1, \sum \limits_{h \in {\cal H}(u):h \neq 0 } x_{f,h} \}$. This establishes condition $1$ of Theorem \ref{Thm:FL}.
    
     Also, the following is true from results in \cite{ageev2004pipage}.    
  \begin{lemma}
       $g \left( \mathbf{X} \right) \geq \left(1 - \left(1 - 1/ d \right)^{d} \right) L \left(  \mathbf{X} \right)$ where $d = \max \limits_{u} \lvert {\cal H}(u)\rvert -1$.
  \end{lemma}   
      \begin{IEEEproof}
       It has been shown in \cite{ageev2004pipage} that for $0 \leq y_i \leq 1$, $1 - \prod \limits_{k=1}^{d} (1- y_k) \geq \left(1-1/d \right)^d \min \{1, \sum \limits_{k} y_k \}$. 
       Applying this to all $g_{f,u} \left( \Xm \right)$ and observing that $(1-(1-1/d)^d)$ is decreasing in $d$ ,we get the above lemma.
      \end{IEEEproof}
  This establishes the lower bound condition in Theorem \ref{Thm:FL}. 
  
Finally, we show the $\epsilon$-convexity condition for $g(\cdot)$. 
  
  \begin{lemma}
      $g \left( \mathbf{X} (\epsilon,\alpha) \right)$ is convex for intermediate $\mathbf{X}(\epsilon,\alpha)$, $\epsilon$ and all possible paths/cycles $\alpha$ that occur in Algorithm \ref{alg:pipageround}.
  \end{lemma}
      \begin{IEEEproof}
        It is sufficient to show that $g_{f,u}(\cdot)$ is convex in $\epsilon$ since $\widetilde{\omega}_u$ and $P_f$ are non negative. Observe that only the variables $x_{f,h}$, for a particular $f$, are involved in the expression for $g_{f,u}(\cdot)$. The edges $(f,h)$ are all incident on a particular vertex $f$. We refer the reader to Algorithm \ref{alg:pipageround} for the definition of the variables used in the proof with $(f,h)$ replacing the edges $(a,b)$. Briefly, $\mathbf{\hat{X}}$ is the current solution at the beginning of any  iteration and $\mathbf{X} \left(\epsilon, \alpha\right)$ is the parametrized solution associated with $\mathbf{\hat{X}}$ during the iteration. Only the variables $\hat{x}_{f,h}$, corresponding to edges participating in $\alpha$, are changed in the iteration.  Since $\alpha$ is either a simple cycle or a simple path, at most two of the variables $x_{f,h} (\epsilon,\alpha)$ are different from $\hat{x}_{f,h}$(by either adding or subtracting $\epsilon$) in any iteration for a given $f$. Also, variables corresponding to one matching are increased and the ones corresponding to the other are decreased. Therefore, if only one variable is changed, then the expression is linear in $\epsilon$ and hence it is convex. If two variables are changed (say $\hat{x}_{f,h_1}$ and $\hat{x}_{f,h_2}$), then either they are changed to $\hat{x}_{f,h_1}+\epsilon$, $\hat{x}_{f,h_2}-\epsilon$ or to $x_{f,h_1}-\epsilon$, $x_{f,h_2}+\epsilon$ . Without loss of generality, assuming one of the cases we have, $g_{f,u} \left( \mathbf{X} (\epsilon,\alpha)\right)= 1-  \left(1-\hat{x}_{f,h_1}-\epsilon \right) \left( 1 - \hat{x}_{f,h_2}+ \epsilon \right) \prod \limits_{h \in {\cal H}(u):h\neq h_1,h_2,0} \left(1-\hat{x}_{f,h} \right)$. This expression is quadratic in $\epsilon$ with upward concavity, and hence it is convex. This proves the theorem. 
 \end{IEEEproof}
 
Now, we apply Theorem \ref{Thm:FL}. 
Consider $\mathbf{R}_{opt}$ to be the optimal solution obtained by maximizing $L=\sum \limits_{f,u} P_f \widetilde{\omega}_u L_{f,u}(\mathbf{R})$ subject to the constraints in 
program (\ref{eqn:IPform2}) where $x_{f,h}$ is replaced by relaxed variables $\rho_{f,h} \in [0,1]$ as follows:

\begin{eqnarray} 
\mbox{maximize} & & \sum_{f=1}^F P_f \sum_{u=1}^{U} \widetilde{\omega}_u \min \{1, \sum \limits_{h \in {\cal H}(u):h \neq 0 } \rho_{f,h} \} \nonumber \\
\mbox{subject to} & & \sum_{f=1}^F \rho_{f,h} \leq M, \;\;\; \forall \;\; h,  \nonumber \\
& & \Rm \in [0,1]^{F \times H}  \label{eqn:LPform}
\end{eqnarray}
 
 Let $\mathbf{X}_{int}$ be the solution obtained by running $\mbox{Pipage Rounding}$$({\cal K}_{\Fc,\Hc},F,M, \mathbf{R}_{opt})$. By Theorem \ref{Thm:FL},  $ g(\Xm_{int}) \geq \left( 1- \left(1-1/d \right)^{d} \right) g (\Xm_{opt})$ where $\Xm_{opt}$ is the optimum to problem (\ref{eqn:IPform2}) and $d = \max_u \{|\Hc(u)|-1\}$ .
 
We note that the terms $\min \{1, \sum \rho_{f,h}\}$ in $L(\cdot)$  can be replaced by variables $t_{f,u}$ with additional constraints $t_{f,u} \leq 1$ and $t_{f,u} \leq \sum \limits_{h \in \Hc(u):h \neq 0} \rho_{f,h} $, in order to turn (\ref{eqn:LPform}) into a linear program (LP) with $(U+H)F$ variables and constraints. Hence, applying the time complexity bound for interior point methods for solving LPs, the algorithm (including pipage rounding) runs with time complexity  $O((U+H)^{3.5} F^{3.5})$. We note that the running complexity of the LP dominates that of the rounding step.
 
 We note two important features of this improved approximation ratio compared to the generic scheme \cite{calinescu2007maximizing} for 
 submodular monotone functions that gives $(1-1/e)$ approximation ratio. First, the generic algorithm runs in time $O (n^8)$ where $n$ is the rank of the matroid. 
 As argued before, in our case, this is typically too complex. 
 Hence, the improved approximation algorithm is faster by orders of magnitude compared to the generic one. 
 Second, in typical practical wireless networks scenarios any user is connected to only a few helpers (e.g.,  $3$ or $4$). 
 The reason is due to spacing between helpers needed to handle interference issues.  
 For example, for the case when every user is connected to at most $4$ helpers, the approximation ratio is $1-(3/4)^{4} \approx 0.6836$ while $1-1/e \approx 0.6321$.  
 Without any constraints on $d$, our result recovers the $1-1/e$ guarantee of the generic algorithm.

%%%%%%%%%%%%%%%%%%%%%%%%%%%%%%%%%%%%%%%%%%%%%%%%%%%%%%%%%%%%%%%%%% 
%%%%%%%%%%%%%%%%%%%%%%%%%%%%%%%%%%%%%%%%%%%%%%%%%%%%%%%%%%%%%%%%%
\section{Coded content placement}

In this section, we consider the optimum cache allocation for the case where the files are encoded by rateless MDS 
coding (e.g., fountain codes \cite{shokrollahi2006raptor}).  With fountain/MDS codes, a file can be retrieved provided that $B$ parity bits are recovered in any order from the helpers or the BS.  Fountain/MDS coding provides a convex relaxation of the uncoded 
problem  studied before. Notice that here we consider only ``intra-session'' coding, i.e., 
we do not consider network coding that mixes bits from different files. 

We let $\Rm = [\rho_{f,h}]$, where $\rho_{f,h}$ denotes the fraction (normalized by $B$) of parity bits of file $f$ contained in the cache of helper $h$. 
The delay to download a fraction of parity bits $\rho_{f,h}$ on the link $(h,u)$ is given by $\rho_{f,h} \omega_{h,u}$. 
A file is entirely retrieved when a fraction larger than or equal to 1 of parity bits is downloaded.
The average delay per information bit necessary for user $u$ to download file $f$, assuming that 
it can download it from its best $j$ helpers, is given by 
 \begin{eqnarray} \label{D(knj)}
\bar{D}_u^{f,j} & =  & \sum_{i=1}^{j-1} \rho_{f, (i)_u} \omega_{(i)_u,u} + \left ( 1 - \sum_{i=1}^{j-1}  \rho_{f, (i)_u} \right ) \omega_{(j)_u,u} \nonumber \\
& = &  \omega_{(j)_u,u}   - \sum_{i=1}^{j-1} \rho_{f, (i)_u} (\omega_{(j)_u,u}  -  \omega_{(i)_u,u}). 
\end{eqnarray} 
Notice that file $f$ can be downloaded by user $u$ from its best $j$ helpers only if
$\sum_{i=1}^j \rho_{f, (i)_u} \geq 1$. In addition, since the BS contains all files, we always have
$\rho_{f,0} = 1$ for all $f \in \Fc$, such that all users can always obtain the requested files by downloading the missing parity bits from the BS. 

The delay $\bar{D}_u^f$ incurred by user $u$ because of downloading file $f$ is a piecewise-defined affine function 
of the elements of the placement matrix $\Rm$, given by 
\begin{equation} \label{piecewise-affine} 
\bar{D}_u^f = \left\{ \begin{array}{ll}
         \bar{D}_u^{f,1} & \mbox{if $\begin{array}{l} \rho_{f,(1)_u} \geq 1 \end{array}$}\\
\vdots & \vdots \\
\bar{D}_u^{f,j} & \mbox{if $\begin{array}{l} \sum_{i=1}^{j-1} \rho_{f,(i)_u} < 1,\\ \sum_{i=1}^j \rho_{f,(i)_u}  \geq 1 \end{array}$}\\
\vdots & \vdots \\
\bar{D}_u^{f,|\Hc(u)|} & \mbox{if  $\sum_{i=1}^{|\Hc(u)|-1} \rho_{f,(i)_u} < 1$}
\end{array} 
\right. 
\end{equation}
We have the following result: 

\begin{lemma}
$\bar{D}_{u}^f$ is a convex function of $\Rm$.
\end{lemma}

\begin{IEEEproof}
A function that is the point wise maximum of a finite number of affine functions is convex \cite{boyd}. Now we show that:
\begin{equation}\label{Dkn}
\bar{D}_{u}^{f} = \max_{j \in \{1,2,\dots,|\Hc(u)|\}} \bar{D}_{u}^{f,j}.
\end{equation}
This means that if for some given $j$, the conditions 
$\sum_{i=1}^{j-1} \rho_{f,(i)_u} < 1$ and $\sum_{i=1}^j \rho_{f,(i)_u}  \geq 1$ hold, 
and therefore $\bar{D}_u^f = \bar{D}_u^{f,j}$, then we have  that 
$\bar{D}_u^{f,j} > \bar{D}_u^{f,j'} \forall \,\, j' \neq j$. 
Thus,  from (\ref{D(knj)}), we should show that:
\begin{align}
\omega_{(j)_u,u}   - \sum_{i=1}^{j-1} \rho_{f, (i)_u} (\omega_{(j)_u,u}  -  \omega_{(i)_u,u})  \geq
\omega_{(j')_u,u}   - \nonumber \\
\sum_{i=1}^{j'-1} \rho_{f, (i)_u} (\omega_{(j')_u,u}  -  \omega_{(i)_u,u})
\end{align}
when the condition for $\bar{D}_u^f = \bar{D}_u^{f,j}$ holds. We only discuss the case $j' > j$ since the proof is similar for the case $j' < j$. 
After some simple algebra, the inequality above can be put in the form
\begin{align}
\left ( \sum_{i=1}^j \rho_{f, (i)_u}   - 1\right ) \left ( \omega_{(j')_u,u} - \omega_{(j)_u,u} \right ) + \nonumber \\
 \sum_{i=j+1}^{j'-1} \rho_{f, (i)_u} \left ( \omega_{(j')_u,u} - \omega_{(i)_u,u} \right )  \geq 0.
\end{align}
This is clearly always verified since  $\sum_{i=1}^j \rho_{f, (i)_u}  \geq  1$ and $\omega_{(j')_u,u} \geq \omega_{(i)_u,u}$ for all $i < j'$. 
\end{IEEEproof}

The average delay of user $u$ is given by  ${{\bar{D}_u}} =\sum_{f=1}^{F}{P_f} \bar{D}_u^f$. 
Using (\ref{Dkn}),  the coded placement optimization problem takes on the form:
\begin{eqnarray} \label{opt_coding}
\mbox{minimize} & &  \sum_{u=1}^{U} \sum_{f=1}^{F}{P_f} \max_{j \in \{1,2,\dots,|\Hc(u)|\}} \left \{ \bar{D}_{u}^{f,j} \right \} \nonumber \\
\mbox{subject to} & &  \sum\limits_{f = 1}^F {{\rho _{f,h}}}  \leq M, \;\;\; \forall \; h \nonumber \\
 & & \Rm \in [0,1]^{F \times H},
\end{eqnarray}
where the optimization is with respect to $\Rm$. Finally, the problem can be reduced to a linear program by introducing the auxiliary 
variables $z_{u,f} \in \RR_+$ and writing (\ref{opt_coding}) in the equivalent form:

\begin{eqnarray}
\mbox{minimize} & &  \sum_{u=1}^{U} \sum_{f=1}^{F}{P_f} z_{u,f} \nonumber \\
\mbox{subject to} & &  z_{u,f} \geq \bar{D}_{u}^{f,j}, \;\;\; \mbox{for} \; j = 1,\ldots, |\Hc(u)|, \;\; \forall \; u,f \nonumber\\
 & & \sum\limits_{f = 1}^F {{\rho _{f,h}}}  \leq M, \;\;\; \forall \; h \nonumber \\
 & & \Rm \in [0,1]^{F \times H}.
\end{eqnarray}
This linear program has $O((U+H)F)$ variables and constraints.  Hence, it has the worst case complexity of $O((U+H)^{3.5}F^{3.5})$ as it has been noted before. In general, optimum value of delay obtained with the coded optimization is better than the uncoded optimization because any placement matrix with integer entries is a feasible solution to the coded problem. In this sense, the coded optimization is a convex relaxation of the 
uncoded problem.

\textbf{Remark}: We further observe that, in the special case where $\omega_{h,u} = \omega_1$ for all $(h,u) \in \Ec$ with $h \neq 0$, and $\omega_1 < \omega_{0,u}$ for all $u \in \Uc$ as in section \ref{sec:specialcase}, $\sum \limits_{u} \omega_{0,u}- \bar{D}_u = L(\mathbf{R})$. Hence, the first LP step of the improved algorithm in section \ref{sec:specialcase}, is equivalent to solving (\ref{opt_coding}) in terms of the final optimal placement matrix $\mathbf{R}$.  In other words, the maximization in problem (\ref{eqn:LPform}) is equivalent to coded minimization in problem (\ref{opt_coding}) for the special case.  Hence, the improved algorithm in section \ref{sec:specialcase} adds a rounding step.

\section{Numerical results} \label{Sec:Numval}

We present numerical results evaluating the greedy uncoded placement and the coded placement in a simplified idealistic setting. We consider 
a cellular region formed by a disk  of radius $350$ m. We assume that the connectivity range of every helper is $70$ m. To compute the various link rates, we assume that the base station operates on a $20$ MHz band with spectral efficiency of $3$ bits/s/Hz while each helper operates on a $20$ MHz band with spectral efficiency of $5$ bits/s/Hz. We ignore interference issues between helpers. As has been noted before, it is a reasonable assumption if neighboring helpers are assumed to operate on orthogonal bands since current $802.11$ WiFi standards allow operation on multiple $20$ MHz bands. 
Assuming that each helper allocates its transmission resource in a fair way,
the rates are calculated by $\mathrm{Spectral ~Efficiency}\times \mathrm{~Bandwidth/Number ~of ~ connected ~ users}$.
In the case of the BS, the number of connected users is $U$ (all users). 
In the case of the helpers, the number of connected users is given by the degree $|\Uc(h)|$ for each helper $h$ in the connectivity 
graph between helpers and users.  We place helpers on a regular grid throughout the system region.
The spacing between grid points can be set in order to produce more or less dense helper positions. 
In all the simulation results, we assume that the size of the entire file library is $F=1000$ and the cache size of each helper 
is $M=100$. For the file popularity distribution $P_f$, we assume a Zipf distribution with parameter $0.56$ (see experimental results in \cite{zink2009}).

\begin{figure}
\centerline{\includegraphics[scale=0.21]{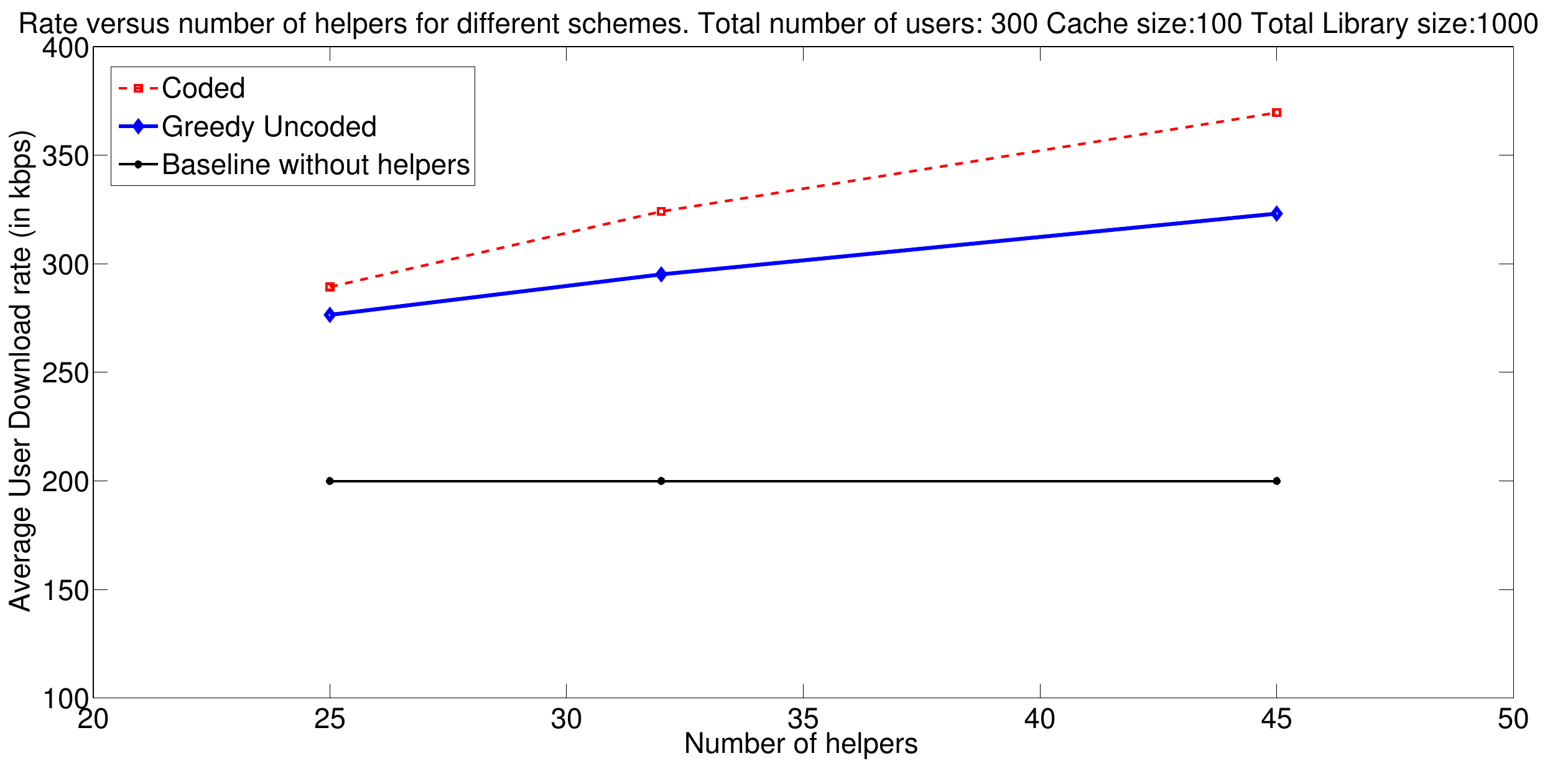}}
\caption{Average user download rate versus number of helpers for $300$ users. }
\label{helpcompare}
\end{figure}

\begin{figure}	
\centerline{\includegraphics[scale=0.21]{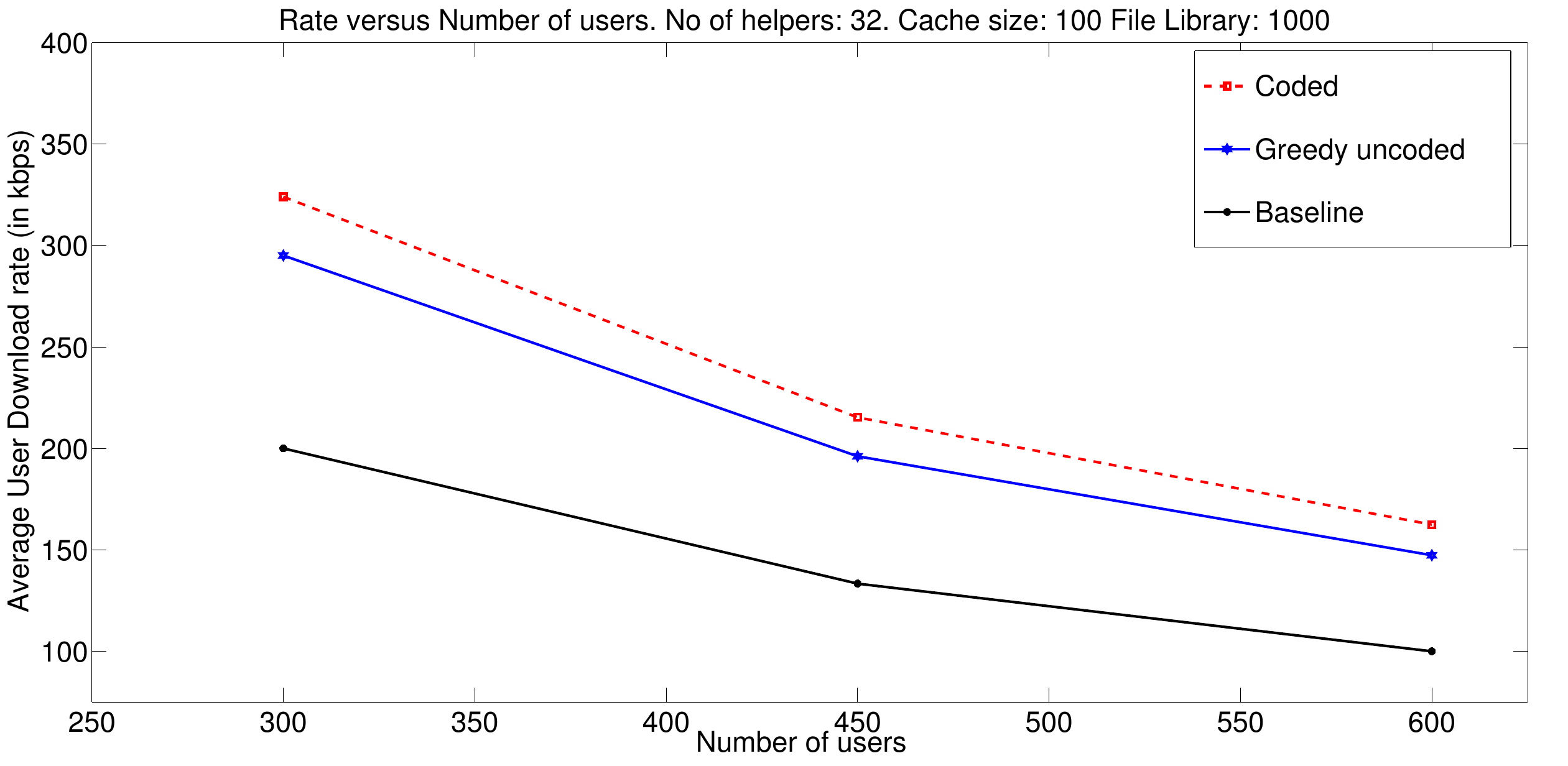}}
\caption{Average user download rate versus number of users for $32$ helpers. }
\label{usercompare}
\end{figure}

\begin{figure}
\centerline{\includegraphics[scale=0.21]{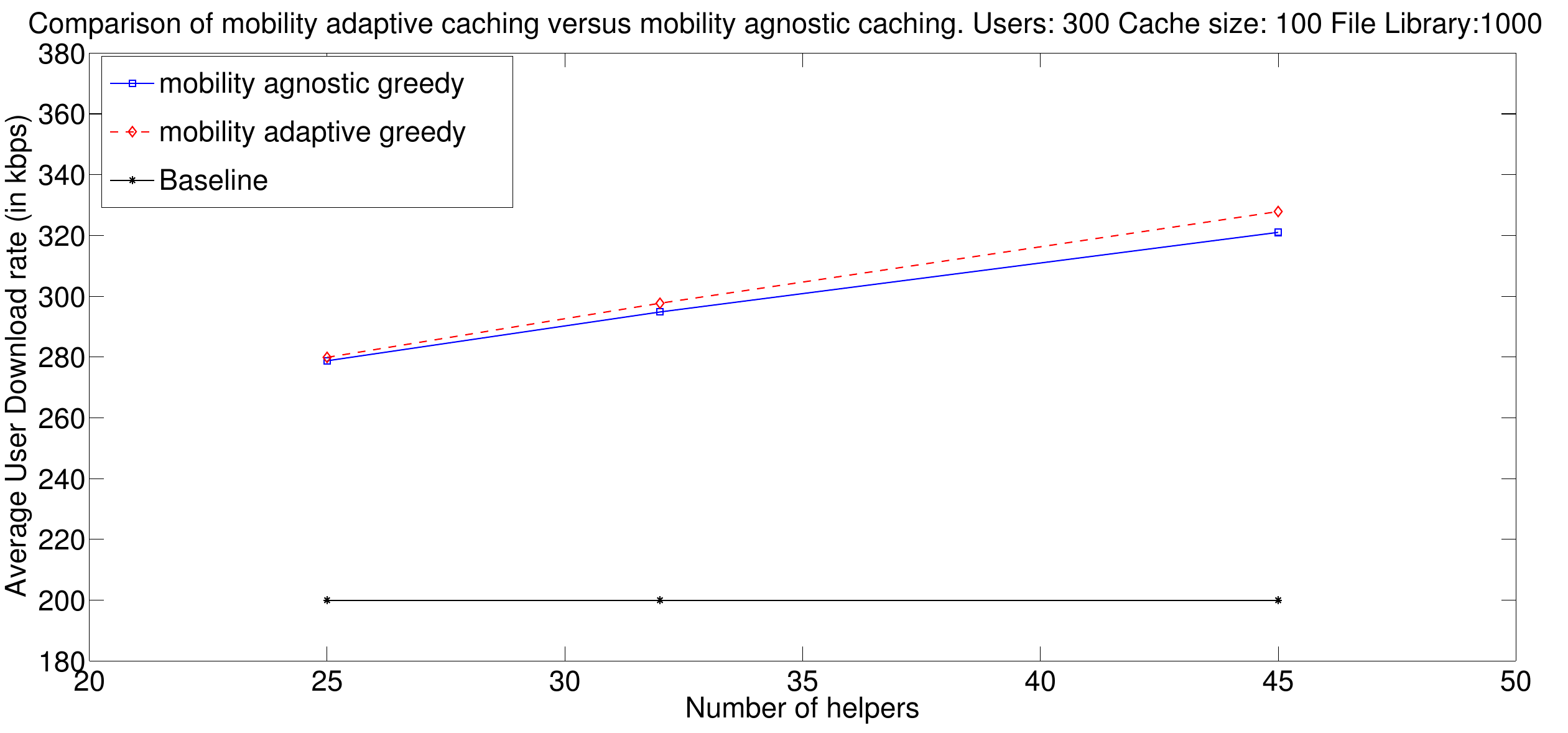}}
\caption{Performance comparison between mobility adaptive and mobility agnostic uncoded placement when users execute independent random walk for $800$ steps each of size $2$ meters. }
\label{mobility}
\end{figure}

For every instance in the simulations, the UT positions are randomly and uniformly generated over the system region and the results are averaged with respect to
the UT positions. We compare three schemes: 1) BS only (without helpers); 
2) BS and helpers with greedy uncoded content placement; 3) BS and helpers with coded placement. 
In Fig. \ref{helpcompare}), the average user download rate under the three schemes is shown for different number of helpers ($H = 25$, $32$ and $45$) 
and $U = 300$ UTs. 
We observe that, as the number of helpers increases, the gap between the performance of the greedy and coded caching schemes increases. 
For the scenario under consideration, we obtain gains of $1.5$x- $2$x with the placement schemes. 
According to the Zipf parameter assumed, the first $100$ files account for $34$ percent of all requests. 
Obviously, the gains are more pronounced when the number of helpers is larger. 
In Fig. \ref{usercompare}, user download rate is plotted under all three schemes for different 
number of users ($U = 300$, $450$ and $600$) in the system and $H = 32$ helpers. 
Finally, Fig. \ref{mobility} deals with the mobility issue that we have ignored in our problem formulation. 
The cache placement algorithms do not take into account the fact that when users are mobile, the connectivity graph between helpers and users changes over time. 
In Fig. \ref{mobility}, we start the system form a random initial condition of the UT positions generated as said before, and then let the users move
according to independent random walks across the system area. 
We consider two cases: 1) adapting the content placement as the connectivity graph changes; 
2) optimizing the placement for the UTs initial positions, and incurring a possible performance degradation as the connectivity graph changes. 
The random walks in our simulation involve at every step independent moves by $2$ meters in one of the principal directions (north, south, east and west) with equal probability, 
and we impose a torus topology at the boundaries such that, when a user gets out of the system on one side, it comes back in on the 
opposite side with respect to the origin.  The content placement is calculated for the initial positions and, for the adaptive case, it is re-calculated after 
$800$ random walk steps for final user positions.  We observe that for this network scenario random user mobility incurs only a small 
performance degradation, which increases as the number of helpers increase. A few remarks are in order. First, it is pleasing to notice that the system performance does not collapse
completely in the presence of low mobility, modeled as users migrating across the helpers in a random way. 
Second, we can explain this fact by noticing that if there is no mass migration with a defined directional drift of the users across the system, 
and if the user density over the system region remains constant over time, all typical user spatial configurations are approximately equivalent 
since the users make requests with the same popularity distribution. Hence, a content placement solution for one configuration is
likely to be quite good also for other typical configurations. This is because the actual identity of the user is irrelevant. If a user moves very far relative to his initial position, qualitatively speaking, there would be another user who would take his initial place. Third, the fact that the gap between optimal placement and fixed placement 
increases with the number of helpers can be explained by noticing that as the helper density increases the number of users per helper with be smaller, 
and therefore the previously mentioned {\em averaging effect} is less relevant. Finally, we hasten to say that these conclusions are expected to hold 
for constant density of the users in the system region and random motion. If all users move to one corner of the region, or migrate outside the region such that 
the density changes, it is obvious that the content placement optimized for the initial network configuration may become arbitrarily bad as the network topology evolves 
over time.

The simulations presented in this paper, ignore link rate changes due to scheduling at the base station. When considering scheduling dynamics at the base station, with caching at the helpers, a big part of the traffic would be offloaded to the helpers due to the placement schemes and hence frequency-time resources available for scheduling are freed to accommodate more demands or support higher downlink rates.
This natural base station `offloading'' advantage of the system with helpers is ignored here and would provide even additional gains to the FemtoCaching 
system with respect to a conventional system with the BS only. 
For a more detailed experimental evaluation, using actual Youtube request traces from a campus \cite{zink2009}, and taking the actual LTE scheduling 
into account in the downlink, we refer the reader to the papers \cite{Infocom,ICC}.

%%%%%%%%%%%%%%%%%%%%%%%%%%%%%%%%%%%%%%%%%%%%%%%%%%%%
\section{Conclusion}

Inspired by the FemtoCaching system proposal, in this paper we focused on
the content placement problem in a wireless network formed by helper nodes and wireless 
users, placing requests to files in a finite library according to a known file popularity distribution. 
%In this paper, we introduced a new method for increasing
%the throughput of wireless video delivery networks. The key
%idea is the use of a distributed cache, \textit{i.e.}, helper stations that
%store the most popular video files, and transmit them, upon
%request, via short-range wireless links to the user terminals.
%The caches are low-cost because storage capacity has become
%exceptionally cheap, while the loading of the files
%to the caches can occur through a low-rate (and thus cheap
%and robust) backhaul links during low demand times. 
We formulated the problem as the minimization of the total expected downloading delay for a given popularity distribution and network topology, 
reflected by the connectivity graph and by the link average rates. 
%and solved the problem of which files should be
%assigned to which helpers assuming that the file popularity distribution and the network connectivity graph and link rates are known. 
We studied two types of placement, namely, coded and uncoded content placement, depending on how the files are stored. 
We showed intractability and developed approximation algorithms for the uncoded scheme.
We also showed that the coded placement is a convex optimization problem. 
Further, we provided numerical results, and numerically addressed the effect of limited mobility 
on the system performance.

For future work, we would like to point out that the uncoded problem is a new coverage problem adding to traditional ones like set cover and 
maximum coverage. It would be actually very interesting to find a better approximation guarantee for the general uncoded problem, 
with improved bounds on the running time.

\section{Acknowledgement}
We thank the anonymous reviewers for helpful comments that helped shape the results presented in \ref{sec:specialcase} and improve the paper in general. 
This work was supported by the Intel/Cisco VAWN program.

%%%%%%%%%%%%%%%%%%%%%%%%%%%%%%%%%%%%%%%%%%%%%%%%%%%%%%%%%%%%%%%%%%%%
%%%%%%%%%%%%%%%%%%%%%%%%%%%%%%%%%%%%%%%%%%%%%%%%%%%%%%%%%%%%%%%%%%%%
%%%%%%%%%%%%%%%%%%%%%%%%%%%%%%%%%%%%%%%%%%%%%%%%%%%%%%%%%%%%%%%%%%%%
%%%%%%%%%%%%%%%%%%%%%%%%%%%%%%%%%%%%%%%%%%%%%%%%%%%%%%%%%%%%%%%%%%%%
\appendices

\section{Basic Definitions}  \label{app:matroids}

\textit{Matroids}:
Matroids are structures that generalize the concept of independence from linear algebra, to general sets. Informally, we need a finite ground set $S$ and a matroid is a way to label some subsets 
of $S$  as ``independent''. In vector spaces, the ground set is a set of vectors, and subsets are called independent if their vectors are linearly independent, 
in the usual linear algebraic sense.  Formally, we have~\cite{combin}:

\begin{definition} \label{matroid-def}
A matroid $\Mc$ is a tuple $\Mc=(S, {\Ic})$, where $S$ is a finite ground set and ${\Ic}\subseteq 2^{S}$ (the power set of $S$) is a collection 
of independent sets, such that:\\
1. {$\Ic$} is nonempty, in particular, $\emptyset \in {\Ic}$,\\
2. {$\Ic$} is downward closed; \textit{i.e.}, if $Y \in {\Ic}$ and $X\subseteq Y$, then $X \in {\Ic}$,\\
3. If $X,Y \in {\Ic}$, and $|X| < |Y|$, then $\exists y \in Y \backslash X$ such that $X \cup \{ y\} \in {\Ic}$. \hfill $\square$
\end{definition}

One example is the partition matroid. In a partition matroid, the ground set $S$ is partitioned into (disjoint) sets $S_{1};S_{2}; ... ;S_{l}$
and 
\begin{equation}\label{partition}
\Ic=\{X\subseteq  S: |X\cap S_{i}|\leq k_{i}\; \mbox{for all} \; i=1\dots l\},
\end{equation}
for some given parameters $k_1,k_2,...,k_l$.

\textit{Submodular functions:}
Let $S$ be  a finite ground set. 
A set function $f : 2^S \rightarrow \mathbb{R}$ is submodular if for all sets $A, B \subseteq S$, 
\begin{equation}
f(A)+f(B) \geq f(A\cup B)+f(A \cap B).
\end{equation}
Equivalently, submodularity can be defined by the following condition. Let 
$f_{A}(i)=f(A+i)-f(A)$ denote the marginal value of 
an element $i \in S$ with respect to a subset $A \subseteq S$. Then, $f$ is submodular if for all $A \subseteq B \subseteq S$ and for all $i \in S\backslash B$ we have:\\
\begin{equation}\label{sub2}
f_{A}(i)\geq f_{B}(i).
\end{equation}
Intuitively, submodular functions capture the concept of diminishing returns: as the set becomes larger the benefit of adding a new element to the set will decrease. 
Submodular functions can also be regarded as functions on the boolean hypercube $\{0,1\}^{|S|} \rightarrow \mathbb{R} $. Every set  has an equivalent boolean representation by assigning $1$ to the elements in the set and $0$ to other ones. We denote the boolean representation of a set $X$ by a vector $X^{b} \in \{0,1\}^{|S|}$.

The function $f$ is monotone if for $A\subseteq B \subseteq S$, we have $f(A) \leq f(B)$.

%\bibliographystyle{IEEEtran}
%\bibliography{infocom_ref,Femtorev}

% NOTE BIBLIOGRAPHY IS GENERATED BY PASTING THE CONTENTS FROM THE .BBL FILE  GENERATED. IF BIBLIPGRAPHY IS UPDATED COMMENT THE FOLLOWING SECTION OUT
%AND USE THE ABOVE TWO LINES THAT REFER TO THE BIB FILES.

\begin{IEEEbiographynophoto}{Karthikeyan Shanmugam}
(S'10) is currently working towards his Ph.D. at the Department of Electrical and Computer Engineering at the University of Texas at Austin. He obtained his B.Tech and M.Tech degrees in Electrical Engineering from the Indian Institute of Technology Madras in 2010 and the M.S. degree in Electrical Engineering from the University of Southern California in 2012. He was an Annenberg Fellow at the University of Southern California for the period between 2010 and 2012. His research interests lie broadly in information theory, coding theory and graph algorithms. His current research focus is on designing and analyzing algorithms for problems in network coding.
 \end{IEEEbiographynophoto}
 
 \begin{IEEEbiographynophoto}
{Negin Golrezaei} received her Bachelor of Science and Master of Science in Electrical Engineering from Sharif University of Technology in 2007 and 2009, respectively. She joined University of Southern California in 2011 with Provost fellowship where she is currently pursuing her Ph.D. degree in Marshall School of Business. Her research interests include information theory, revenue management, mechanism design and 
optimization algorithms.
\end{IEEEbiographynophoto}

\begin{IEEEbiographynophoto}{Alexandros G. Dimakis} (S'01 -- M'09)
Alex Dimakis is an Assistant Professor at the Electrical and Computer Engineering department, University of Texas at Austin. From 2009 until 2012 he was with the Viterbi School of Engineering, University of Southern California. He received his Ph.D. in 2008 and M.S. degree in 2005 in electrical engineering and computer sciences from UC Berkeley and the Diploma degree from the National Technical University of Athens in 2003. During 2009 he was a CMI postdoctoral scholar at Caltech.

He received an NSF Career award in 2011, a Google faculty research award in 2012 and the Eli Jury dissertation award in 2008. He is the co-recipient of several best paper awards including the joint Information Theory and Communications Society Best Paper Award in 2012. He is currently serving as an associate editor for IEEE Signal Processing letters. His research interests include information theory, signal processing, and networking, with a current focus on distributed storage and machine learning.
\end{IEEEbiographynophoto}

\begin{IEEEbiographynophoto}
{Andreas F. Molisch} (SÕ89 -Ð MÕ95 Ð- SMÕ00 Ð-FÕ05) received the Dipl. Ing., Ph.D., and habilitation degrees from the Technical University of Vienna, Vienna, Austria, in 1990, 1994, and 1999, respectively. He subsequently was with AT\&T (Bell) Laboratories Research (USA); Lund University, Lund, Sweden, and Mitsubishi Electric Research Labs (USA). He is now a Professor of electrical engineering with the University of Southern California, Los Angeles.

His current research interests are the measurement and modeling of mobile radio channels, ultra-wideband communications and localization, cooperative communications, multiple-inputÐmultiple-output systems, wireless systems for healthcare, and novel cellular architectures. He has authored, coauthored, or edited four books (among them the textbook Wireless Communications, Wiley-IEEE Press), 16 book chapters, some 160  journal papers, and numerous conference contributions, as well as more than 70 patents and 60 standards contributions.

Dr. Molisch has been an Editor of a number of journals and special issues, General Chair, Tecnical Program Committee Chair, or Symposium Chair of multiple international conferences, as well as Chairman of various international standardization groups. He is a Fellow of the IEEE, Fellow of the AAAS, Fellow of the IET, an IEEE Distinguished Lecturer, and a member of the Austrian Academy of Sciences. He has received numerous awards, among them the Donald Fink Prize of the IEEE, and the Eric Sumner Award of the IEEE.
\end{IEEEbiographynophoto}

\begin{IEEEbiographynophoto}{Giuseppe Caire}
(S'92 -- M'94 -- SM'03 -- F'05) was born in Torino, Italy, in 1965. He received the B.Sc. in Electrical Engineering  from Politecnico di Torino (Italy), in 1990, 
the M.Sc. in Electrical Engineering from Princeton University in 1992 
and the Ph.D. from Politecnico di Torino in 1994.   He was a recipient of the AEI G.Someda Scholarship in 1991,
has been with the European Space Agency (ESTEC, Noordwijk, The Netherlands) from May 1994 to February 1995, 
was a recipient of the COTRAO Scholarship in 1996 and of a CNR Scholarship in 1997. 

He has been visiting Princeton University in Summer 1997 and Sydney University  in Summer 2000.  He has been Assistant Professor in Telecommunications at the Politecnico di Torino, Associate Professor at the University of Parma, Italy,  Professor with the Department of Mobile Communications at the Eurecom Institute,  Sophia-Antipolis, France, and he is currently a professor of Electrical Engineering with the Viterbi School of Engineering, University of Southern California, Los Angeles, CA.

He served as Associate Editor for the IEEE Transactions on Communications in 1998-2001 and as Associate Editor for the IEEE Transactions on Information Theory in 2001-2003.  He received the Jack Neubauer Best System Paper Award from the IEEE Vehicular Technology Society in 2003,  and the
IEEE Communications Society \& Information Theory Society Joint Paper Award in 2004 and in 2011. 
Giuseppe Caire has served in the Board of Governors of the IEEE Information Theory Society from 2004 to 2007,
and as President of the  IEEE Information Theory Society in 2011. 
His main research interests are in the field of communications theory, information theory, channel and source coding
with particular focus on wireless communications.   
\end{IEEEbiographynophoto}

\end{document}